\documentclass[12pt,twoside]{article}
\usepackage{fleqn,espcrc1}
\usepackage{graphicx}
\usepackage[figuresright]{rotating}
\newcommand{\slsh}[1]{\slash \!\!\! #1}

\newcommand{\AmS}{{\protect\the\textfont2
  A\kern-.1667em\lower.5ex\hbox{M}\kern-.125emS}}
\title{
\vspace*{-2.8cm}
\hfill{\small {MKPH-T-00-11}}\\
\hfill{\small { TRI-PP-00-31}}\\[1.4cm]
A simple model illustrating the impossibility of measuring
off-shell effects}
\author{S. Scherer\address{Institut f\"{u}r Kernphysik, Johannes
Gutenberg-Universit\"{a}t, 55099 Mainz, Germany}
        and 
H.W. Fearing\address{TRIUMF, 4004 Wesbrook Mall, Vancouver, British Columbia,
Canada V6T 2A3}}
\begin{document}
\maketitle
\begin{abstract}
   We consider a simple model and make use of field transformations
in Lagrangian field theories to illustrate the impossibility of measuring
off-shell effects in nucleon-nucleon bremsstrahlung
and related processes.

\end{abstract}

\section{INTRODUCTION}

   The issue of how to describe the interaction of a particle which is 
not on its mass shell has a long history.
   For example, in their derivation of the low-energy theorem of Compton
scattering in 1954, Gell-Mann and Goldberger \cite{GellMann_1954}
took into account that the electromagnetic vertex of an off-shell nucleon is 
more complicated than that of a free nucleon.
   In case of the electromagnetic interaction, some restrictions on the general
form of the off-shell vertex result from the Ward-Takahashi identity
\cite{WT}.
   It is a natural and legitimate question to ask whether it is possible 
to extract the off-shell behavior of particular interaction vertices from 
empirical information similarly as one, say, extracts the electromagnetic
form factors of the nucleon from elastic electron scattering.
   In this context one might think of the electromagnetic interaction of
a bound nucleon \cite{BN}. 
   Another example would be an investigation of the off-shell nucleon-nucleon
amplitude entering the nucleon-nucleon-bremsstrahlung process
\cite{Theory}.

   It has only been recently that the relevance of field transformations
in the framework of Lagrangian field theories \cite{FT}
has been emphasized in addressing this question.    
   For the case of pions, Compton scattering \cite{Scherer_1995} and pion-pion
bremsstrahlung \cite{Fearing_1998} have been considered using
chiral perturbation theory.
   It was shown that off-shell effects both depend on
the model used and on the choice of representation for the fields.
   From that we concluded that off-shell effects are not only
model dependent but also representation dependent, making
a unique extraction of off-shell effects impossible.
   In practice, the freedom of choosing appropriate field variables has been 
used in Refs.\ \cite{LET} to obtain the
most general effective Lagrangian describing low-energy
(virtual) Compton scattering.

   Here, we will extend our previous discussion 
\cite{Scherer_1995,Fearing_1998} to the investigation of
a spin one-half system \cite{Fearing_2000}
which should remove any uncertainty that the results 
of the previous works somehow depended on the simplicity of a spin-zero 
process.

\section{THE MODEL}
   As our toy model we take 
\begin{equation}
{\cal L}_0 = \overline{\Psi}(i D \hspace{-.6em}/ -m) \Psi -\frac{e \kappa}{4 m}
\overline{\Psi} \sigma_{\mu \nu} F^{\mu \nu} \Psi
+\overline{\Phi}(i \partial\hspace{-.5em}/ -m)\Phi
+g \overline{\Psi} \Psi
\overline{\Phi} \Phi,
\label{eq:L0}
\end{equation}
   where $D_\mu \Psi = (\partial_\mu +ieA_\mu)\Psi$ is the covariant 
derivative of the proton field, $e$ and $\kappa$ are the proton charge 
and anomalous magnetic moment respectively, $A_\mu$ is the photon field, 
and $F_{\mu\nu} = \partial_\mu A_\nu - \partial_\nu
A_\mu$ is the electromagnetic field strength tensor.
   The fields $\Psi$ and $\Phi$ refer to protons and neutrons, respectively.
   For calculational simplicity, we neglect the electromagnetic coupling to 
the neutron magnetic moment.
   In the framework of Eq.\ (\ref{eq:L0}) it is straightforward to calculate 
the Born diagrams of proton-neutron bremsstrahlung:
\begin{eqnarray}
{\cal M}_0^{NN\gamma}&=&
ieg \overline{u}_n(p_4)u_n(p_2) \overline{u}_p(p_3)\left[ \left(
\slsh{\epsilon} -\frac{i \kappa}{2m} \sigma_{\mu\nu} \epsilon^\mu k^\nu \right)
\frac{\slsh{p_3}+\slsh{k} +m}{(p_3+k)^2 -m^2} \right. \nonumber \\ &+& \left.
\frac{\slsh{p_1}-\slsh{k} +m}{(p_1-k)^2 -m^2}
\left( \slsh{\epsilon} -\frac{i \kappa}{2m} \sigma_{\mu\nu}
\epsilon^\mu k^\nu \right) \right] u_p(p_1),
\label{eq:nngborn}
\end{eqnarray}
   which corresponds to the usual choice for nucleon-nucleon
bremsstrahlung for the electromagnetic parts.
   Obviously, for pedagogical reasons, Eq.\ (\ref{eq:nngborn}) has a much 
simplified interaction for the strong part.

\section{FIELD TRANSFORMATIONS}
   Next, we perform a change of variables
leading to different off-shell behavior in the nucleon-nucleon amplitude as
well as the photon-nucleon vertex, and study the effect on the total
$pn$ bremsstrahlung amplitude.
   For that purpose we consider the transformation 
\begin{equation}
\Psi \rightarrow \Psi + \tilde{a}g \overline{\Phi}\Phi\Psi
+\tilde{b}e \sigma_{\mu\nu} F^{\mu\nu} \Psi,
\label{eq:wftrans}
\end{equation}
   where $\tilde{a}$ and $\tilde{b}$ are real constants,  determining the 
overall strength of the transformation.
   Equation (\ref{eq:wftrans}) results in a class of
equivalent Lagrangians
\begin{equation}
{\cal L}(\tilde{a},\tilde{b})={\cal L}_0+\Delta {\cal L}(\tilde{a},\tilde{b})
={\cal L}_0+\Delta {\cal L}_1(\tilde{a},\tilde{b}) + \Delta
{\cal L}_2(\tilde{a},\tilde{b}),
\label{dl}
\end{equation}
   where the explicit expressions of $\Delta{\cal L}_1$ and $\Delta{\cal L}_2$
can be found in Ref.\ \cite{Fearing_2000}.    
   Consider, for example, $\tilde{a}\neq 0$ and $\tilde{b}=0$, in which case 
the $\tilde{a}$ term will generate an off-shell dependence
in the proton legs of the $pn$ amplitude
\begin{equation}
i \tilde{a}g [(\slsh{p}_3 -m) + (\slsh{p}_1 -m)] 
\overline{u}_n(p_4)u_n(p_2),
\label{eq:L1strongvertex}
\end{equation}
in conjunction with a $pn\gamma$ contact term
\begin{equation}
-2 i \tilde{a}eg \overline{u}_p(p_3) \left(
\slsh{\epsilon} -\frac{i \kappa}{2m} \sigma_{\mu\nu} \epsilon^\mu k^\nu 
\right) u_p(p_1)
\overline{u}_n(p_4)u_n(p_2).
\label{eq:L1acontact}
\end{equation}
   One can show that in the amplitude the off-shell dependence 
from Eq.\ (\ref{eq:L1strongvertex}) precisely cancels the contact
term of Eq.\ (\ref{eq:L1acontact}).
   In the present context, such a cancellation for the charge, but not the
magnetic, parts is simultaneously 
enforced by gauge invariance. 
   A similar observation was also made in Ref.\ \cite{Kaloshin_1999}
for real Compton scattering off a charged pion.
   However, the implication of field transformations is even more general,
because it also has consequences for terms which are not fixed by gauge
invariance, such as the magnetic terms above. 
   The general case $\tilde{a}\neq 0\neq \tilde{b}$ is discussed in Ref.\
\cite{Fearing_2000}.

   The above result is a simple illustration of the equivalence theorem of 
Lagrangian field theory \cite{FT}, according to which all Lagrangians of 
Eq.\ (\ref{dl}) result in identical $S$-matrix elements.
   One can also make use of this observation in order to show that what 
appears as an off-shell effect in an $S$-matrix 
element for one Lagrangian may originate
in a contact term from an equivalent Lagrangian \cite{Fearing_2000}.

\section{CONCLUSION}

   We conclude that off-shell effects cannot in any unambiguous way be 
extracted from an $S$-matrix element.

\end{document}